\documentclass[aps,amsfonts,nofootinbib,superscriptaddress,twocolumn]{revtex4-1}

\usepackage{amsmath,amssymb}

\usepackage{appendix}
\usepackage{dsfont}
\usepackage{amssymb}
\usepackage{amsmath}
\usepackage{amsbsy}
\usepackage{color}
\usepackage{slashed}
\usepackage{bm}
\usepackage{bbm}
\usepackage{dutchcal}
\usepackage{tikz}
\usepackage{xparse}
\usepackage[normalem]{ulem}

\usetikzlibrary{calc}
\usetikzlibrary{positioning,arrows,patterns,automata}
\usetikzlibrary{decorations.pathmorphing,decorations.markings,trees}
\usetikzlibrary{shapes.misc}

\tikzset{	photon/.style={decorate, decoration={snake}, draw=black},
		particle/.style={draw=black, postaction={decorate},
        			decoration={markings,mark=at position .5 with {\arrow[draw=black]{>}}}},
	    	aparticle/.style={draw=black, postaction={decorate},
        			decoration={markings,mark=at position .5 with {\arrow[draw=black]{<}}}},
	    	axion/.style={draw=black},
		gluon/.style={decorate, draw=red,
        			decoration={coil,amplitude=4pt, segment length=5pt}},
		vertex/.style={draw,shape=circle,fill=black,minimum size=80pt,inner sep=1pt},
		cross/.style={cross out, draw=black, minimum size=2*(#1-\pgflinewidth), inner sep=0pt, outer sep=0pt}, 				cross/.default={0.3cm}}
		
 \NewDocumentCommand\semiloop{O{black}mmmO{}O{above}}
{\draw[#1] let \p1 = ($(#3)-(#2)$) in (#3) arc (#4:({#4+180}):({0.5*veclen(\x1,\y1)})node[midway, #6] {#5};)}

\newcommand{\ee}{\end{equation}}
\newcommand{\bb}{\begin{equation}}
\newcommand{\eqb}{\begin{eqnarray}}
\newcommand{\eqf}{\end{eqnarray}}

\def\sigmavec{\mbox{\boldmath$\sigma$}}
\def\nablavec{\mbox{\boldmath$\nabla$}}

\def\L{{\bf L}}
\def\J{{\bf J}}
\def\bx{{\boldsymbol \xi}}
\begin{document}
\title{ Anapole Dark Matter Quantum Mechanics }

 \author{
 J. Gamboa 
 }
\email{jorge.gamboa@usach.cl}
\affiliation{Departamento de  F\'{\i}sica, Universidad de  Santiago de
  Chile, Casilla 307, Santiago, Chile}
\author{
 F. M\'endez
 }
\email{fernando.mendez@usach.cl}
\affiliation{Departamento de  F\'{\i}sica, Universidad de  Santiago de
  Chile, Casilla 307, Santiago, Chile} 
\author{
 N. Tapia
 }
\email{natalia.tapia@usach.cl}
\affiliation{Department of Physics, Virginia Tech, Blacksburg VA 24061, U.S.A}  
    \date{\today}
\begin{abstract} 
The dynamics of an anapole seen as dark matter at low energies is studied by solving the Schr\"odinger-Pauli equation in a potential involving Dirac-delta and its derivatives in three-dimensions. 
This is an interesting mathematical problem that, as far as we know, has not been previously discussed. We show how bound states emerge in 
this approach and the scattering problem is formulated (and solved) directly. 
The total cross section is in full agreement with independent calculations in the standard model.\end{abstract}
\date{\today}
\maketitle

\section{Introduction}

In the context of dark matter it is widely believed that some discrete symmetries are not conserved, 
in the same way that in the visible sector, for example,  the parity is broken in the weak interactions. 
In both cases -- dark and visible matter -- if parity is violated an anapole term emerges as the  result
of the interaction between the spin of the fermions and an electromagnetic source. 

This last fact is  obtained taking the non-relativistics limit of
\bb 
{\cal L} =  {\bar \chi} \left( i {   {{ \partial }  \hspace{-.6em}  \slash
      \hspace{.15em}}}    -{\bar m} \right) \chi  -\frac{g}{M^2} {\bar \chi}\gamma_5 \gamma_\mu \chi \partial_\nu F^{\mu \nu},
       \label{correct1}
\ee
 where $M$ is a mass scale, $\bar{m}$ is the mass of  fermions $\chi$ (electrically neutral), $\gamma^\mu$ 
 are the Dirac matrices  (with $\mu =0,\cdots, 3$),  $F^{\mu\nu}$ is the electromagnetic strength tensor and
 $g$ is the dimensionless  coupling constant.
 
 The interaction term in the Lagrangian (\ref{correct1}) was proposed by Zeldovich \cite{zel} sixty years ago as a way to produce
photons by neutral particles such as neutrinos, and extending the ideas of parity violation previously proposed
by Lee and Yang \cite{yang-lee}. 
 
 This term was named by Zeldovich himself as an
{\it anapole} interaction because it is the natural extension of the 
multipolar expansion for truly neutral interactions (for a
review see \cite{review}). 
 
 Even more, in the context of Lee and Yang's parity violation \cite{yang-lee}, Zeldovich pointed out \cite{zel} that the process $ \nu  {\bar \nu}$ could also  produce a virtual photon and therefore violate parity  assuming an effective vertex as in the Figure 1. Since the incoming 
{particles} are neutral, the interaction is { described through the} anapole which is,  technically, {  the analog  of the next 
to quadrupolar term in the } multipolar expansion of 
$ \frac{1}{|{\bf x} -{\bf y}|}$  in electrodynamics. However, although this effect is very weak, it is measurable and 
its detection was announced  in 1997    \cite{wood}  for the transition  $6S$ to $7S$ in cesium atoms  (see also \cite{flaum}).
 
 The idea of anapoles  has been used recently by several authors \cite{mas} and mainly by 
 Ho and Schrerrer \cite{scherrer} who have proposed that the anapole 
 could be considered a form of dark matter. 
 
{{ In this paper we study some properties of this anapole interaction
 from the point of view of quantum mechanics and we show the emergence of  non-trivial properties 
  such as the formation of bound states and a notorious simplicity in scattering processes.}} 
 
 The paper is organized as follows: in section II we discuss the non-relativistic anapole and we formulate the problem in general, in section III we discuss the bound state and scattering problem in three dimensions and we provide of exact solution of the Sch\"odinger-Pauli equation, in section IV we discuss our results. Two appendices of technical issues are included. 
 
 \section{Non-Relativistic Anapole}
 
 The non-relativistic  { limit  \cite{nos}  of the Hamiltonian} (\ref{correct1}) is
\eqb 
H&=& \frac{1}{2m} {\bf p}^2 - \frac{g}{M^2} {\bf S} \cdot {\bf J} + {\cal O} (c^2) \nonumber 
\\ 
&=&   \frac{1}{2m} {\bf p}^2 +   H_{\mbox{\tiny{anap}}} + {\cal O} (c^2), \label{correct2}
\eqf  
where the first term is the kinetic energy of the  leptons with three-momentum ${\bf p}$ 
and $H_{\mbox{{\tiny anap}}}= -  \frac{g}{M^2}{\bf S} \cdot {\bf J}$ is the { Hamiltonian describing } the anapole interaction,  
 { with } ${\bf S} =\frac{1}{2}\sigmavec$ its spin and ${\bf J}$ is the electromagnetic source (coming from the Ampere's law). 

In the case of dark matter  the above issue is a little bit more involved because the interaction between visible and dark matter 
is indirect and the concept of kinetic mixing is necessary \cite{kinetic}. 

The interaction between the dark fermions and the gauge fields is produced by the diagonalization of the kinetic mixing implying that instead of (\ref{correct2}), for the case of dark matter one {has to consider}
\eqb 
H&=& \frac{1}{2m} {\bf p}^2 - \frac{g'}{M^2} {\bf S}_{\mbox{\tiny {dark}}} \cdot {\bf J} + {\cal O} (c^2) \nonumber
\\
&=& \frac{1}{2m} {\bf p}^2 +  H_{\mbox{\tiny dark}} + {\cal O} (c^2), \label{correct3}
\eqf 
where,  now,  $g'=\xi g$ is an effective coupling constant, $\xi\ll 1$ is the kinetic mixing parameter and the anapole  describes the interaction between dark matter and the electromagnetic source. 

The change $g\to g'$  is a nontrivial  consequence of the diagonalization of the kinetic 
mixing and  the gauge group { enlargement} $U (1)$ to $U (1)\times U (1)$.

Then as ${\bf J}$ is a current density of { particles of } the standard model,  we can assume that the particles are millicharged and the  
Ohm's law implies ${\bf J} = {\rho}~ {\bf v} $ where ${\bf  v} $ is the velocity of the charge carriers moving in a volume of $V$ with the charge density $\rho$. 

As the charge carriers are point particles $ \rho \sim \delta ({\bf x})$,and  classical anapole Hamiltonian can be written as
\bb
H_{\mbox{\tiny anap}} = -\frac{g'}{M^2 m} ~\sigmavec \cdot{\bf p} ~\delta({\bf x}), 
\ee
which is an \lq \lq electromagnetic" interaction $ {\bf J}\cdot{\bf A} $ where $ {\bf A }$ is formally identified with $\sigmavec \delta ({\bf x}) $ which is a  Aharonov-Bohm-like effect.  In other words, we have a toroidal cylinder with a magnetic field confined inside in a similar way to the Tonomura experiments for the Aharonov-Bohm effect \cite{tono}.

Then,   in our non-relativistic quantum mechanics model, particles with momentum ${\bf p} $ interact with dark matter through an  
anapole term, giving rise to the total Hamiltonian operator is  
\bb 
{\hat H}= \frac{1}{2m} {\bf p}^2 - \frac{g'}{2M^2 m} \left(  \delta ({\bf x}) ~\sigmavec \cdot {\bf p}+\sigmavec \cdot {\bf p} ~\delta ({\bf x}) \right), \label{sola2}
\ee 
and the solution to this problem will provide a complete quantum mechanical picture, including bound and scattering states.

 
 In order to solve { the eigenvalue}  problem, let us start considering the following Schr\"odinger-Pauli equation
 \eqb
 \left(\nablavec^2 +k^2 \right) \psi ({\bf x}) &=& -\frac{i g'}{M^2 } \bigl[  \delta ({\bf x}) ~\sigmavec \cdot \nablavec \psi ({\bf x})  \nonumber 
 \\
 &+&\sigmavec \cdot \nablavec ~\delta ({\bf x}) \psi ({\bf x}) \bigr],
 \label{sch1}
 \eqf
 where $k^2 = 2 \,m\,E$. 
 
{ Previous equation (\ref{sch1}), can be recast as an integral equation}\footnote{By convenience we will do the calculation in $D$ dimensions.}
 \eqb 
&& \psi ({\bf x}) = \psi_0 ({\bf x}) -\frac{i g'}{M^2 } 
 \int d^D x' G[{\bf x},{\bf x}'] \bigl[ \delta ({\bf x}')~\sigmavec\cdot\nablavec' \psi({\bf x}')  \nonumber 
 \\
 &+& \sigmavec \cdot \nablavec' \delta ({\bf x}') \psi({\bf x}') \bigr]
 \label{sch01}
 \\
 &=& \psi_0 ({\bf x}) -\frac{i g'}{M^2 } \big( G[{\bf x}, 0] \sigmavec \cdot \nablavec \psi (0) - \sigmavec \cdot \nablavec G[{\bf x},0] \psi (0)\bigr), \nonumber
 \\
  \label{sch12}
 \eqf
with  $\psi_0 ({\bf x}) = A~e^{\imath{\bf k}\cdot{\bf x}}$,  the homogeneous solution of the operator $\nablavec^2+k^2$,
and  the Green function $G[{\bf x}, {\bf x}']$ given by
\eqb
G[{\bf x}, {\bf x}'] &=& \int \frac{d^Dp}{(2 \pi)^D} \frac{e^{i {\bf p}\cdot({{\bf x}-{\bf x}'})}}{{\bf p}^2 + k^2} \nonumber
\\
&=&  \left(\frac{|{\bf x}-{\bf x}'|}{2 |k|}\right)^{1-\frac{D}{2}} K_{1-\frac{D}{2}}\left( |k|~|{\bf x}-{\bf x}'|\right).
\label{greend}
\eqf

An instructive example is  to consider the one-dimensional  case, where  previous solution reduces to
\bb
 \psi ({ x}) =  \psi_0 ({x}) -\frac{i g'}{M^2 } \left[  G[x,0]   \psi' (0) - G'[x,0] \psi (0)\right],
  \label{sch13}
 \ee
{ It is interesting to determine the bound states of this problem. Then $\psi_0 = 0$  and and we evaluate
previous expression at   $x =0$, that is } 
\bb
\left( 1- \frac{i g'}{M^2 } G'[0,0] \right) \psi(0) = -\frac{i g'}{M^2 } G[0,0] \psi'(0). 
\ee
We impose Robin's boundary condition 
\bb 
\psi' (0) = \gamma \psi(0),
 \label{robin}
\ee
where $\gamma \in \Re$ is the parameter that defines the self-adjoint extension of the Hamiltonian, to 
find 
\bb 
\left(1-\frac{ig'}{M^2} G'[0,0]\right) = -\frac{ig'\gamma G[0,0]}{M^2}.
\ee
The functions $G[0,0]$ and $G'[0,0]$, from (\ref{greend}), are
\bb 
G[0,0]= -\frac{i\pi}{k}, ~~~~~~~~~
G'[0,0] =0, 
\ee 
and therefore
\bb
E= \frac{k^2}{2m} = \frac{{g'}^2\gamma^2 \pi^2}{2m M^4} >0, 
\ee 
and therefore there are not bounds states in the one-dimensional case.

%
\section{Three-Dimensional Non-Relativistics Anapole; Bound and Scattering States}

The three-dimensional case is more complicated and has important differences with the previous one-dimensional example, such as the existence of bound state and renormalization of the coupling constant as a consequence of the three-dimensional scale invariance \cite{jackiw,garcia} \footnote{By continuity of the presentation of our results, some technical and notation aspects are relegated to Appendices A and B.}.

Then for bound states we put $\psi_0=0$ and (\ref{sch01}) becomes 
\eqb  
\psi ({\bf x}) &=& -\frac{ig'}{2M^2} \int d^3 x' G[{\bf x},{\bf x}'] \delta ({\bf x}')~\sigmavec\cdot\nablavec' \psi({\bf x}') \nonumber
\\
&-& \frac{ig'}{2M^2} \int d^3 x' G[{\bf x},{\bf x}'] ~\sigmavec\cdot\nablavec' \left(\delta ({\bf x}')~ \psi({\bf x}') \right). ~\label{tree}
\eqf

We are interested in the case ${\bf x} =0={\bf x}'$ and due to the   spherical symmetry  and 
the explicit form of the  Green function one  gets $\partial_r G\to 0$. 

Therefore we write (\ref{tree}) as follows
\begin{equation}
\label{solscatt}
 \psi({\bf x}) = -\frac{\imath g'}{M^2}\,\int G({\bf x},{\bf x}')
 \bigg[\delta({\bf x}')({\boldsymbol \sigma}\cdot{\boldsymbol  \nabla}')\bigg]\psi({\bf x}')\,d^3x'.
\end{equation}

Integral (\ref{solscatt})  can be done straightforwardly. To do that, we write explicitly the spinors and look for solutions with the form
\begin{equation}
\label{sloantz}
\psi({\bf x}) = \Phi_+\,\bx^+ + \Phi_-\,\bx^-,
\end{equation}
where $\Phi^{\pm}$ are functions of $r$, while spinors carry the angular dependence.  

It is interesting to note that in usual cases the radial functions are equals ($\Phi_+ =\Phi_-$), but in our case the interaction changes the orbital angular momentum  states according to (\ref{actspin}), what support our choice and then the state
is a superposition of the two different orbital angular momentum $\ell=j\pm\frac12$ for a fixed total angular momentum $j$.

Now we use the explicit form of the Green function and integrate (\ref{solscatt}) for the solution in (\ref{sloantz})
and we evaluate in ${\bf x}=0$.  
One gets
\begin{equation}
\bigg[\Phi_+ (0) - \frac{\imath g'}{M^2}\,G_\Lambda \,\Phi_-'(0)\bigg]\bx^+ =
\bigg[\Phi_-(0) - \frac{\imath g'}{M^2}\,G_\Lambda\,\Phi_+'\bigg]\bx^-,
\end{equation}
where $g'$ has been rescaled through $g'\to g'(1+\theta(0))$ as a consequence of scale invariance \cite{jackiw,garcia}. 

The Green function $G_\Lambda$, instead, has been regularized 
through an ultraviolet cutoff $\Lambda$ for ${\bf x},{\bf x}'\to0$ (see appendix B).

Previous equations can be recast as
\begin{equation}
\left( \begin{array}{c}
\Phi_+
\\
\Phi_-
\end{array}\right)_0
=\frac{\imath g'\,G_\Lambda}{M^2} \left(
\begin{array}{cc}
0 & 1
\\
1 & 0
\end{array}\right)
\left( \begin{array}{c}
\Phi_+'
\\
\Phi_-'
\end{array}\right)_0,
\end{equation}
where  subscript $0$  stand for  ${\bf x}=0$. 

In order to find the bound states we posit a generalization of the  Robin's 
boundary conditions, as follows
\begin{equation}
\label{cond1}
\left( \begin{array}{c}
\Phi_+'
\\
\Phi_-'
\end{array}\right)_0 = \mathbb{{G}}\left( \begin{array}{c}
\Phi_+
\\
\Phi_-
\end{array}\right)_0,
\end{equation}
with $\mathbb{G}$  a  $2\times2$ matrix {{with dimensions of mass}}.

With this choice, condition (\ref{cond1}) read 
\begin{equation}
\left(\mathbb{I} - \frac{\imath g'\,G_\Lambda}{M^2}\,\sigma_1{\mathbb G}\right)\left( \begin{array}{c}
\Phi_+
\\
\Phi_-
\end{array}\right)_0
=0,
\end{equation}
and the condition for bound sates is 
\begin{equation}
\label{cond2}
\mbox{det}
\left(\mathbb{I} - \frac{\imath g'\,G_\Lambda}{M^2}\,\sigma_1{\mathbb G}\right)
=0.
\end{equation}

We look for imaginary solutions for $k$, since then $k^2<0$,  {corresponding to  negative energy states}.
For the Green function (see appendix B)
$$
G_\Lambda = \frac{k}{4\pi\imath}-\Lambda,
$$
we choose
$$
{\mathbb G} = {{\kappa}} \left(
\begin{array}{cc}
\alpha & -\imath\, \beta
\\
\imath\,\beta& -\alpha
\end{array}
\right),
$$
with $\{\alpha,\beta\}\in\Re$, $\alpha^2+\beta^2=1$  {{and $\kappa$, a constant with  dimensions of mass.  Note that }}$\mathbb{G}^\dag\,\mathbb{G}=\kappa^2\,\mathbb{I}$.
By doing this, we get 
\begin{equation}
k_\pm=4\pi\imath\left( \Lambda \pm \frac{M^2}{\kappa\,g' } \right),
\end{equation}
and, therefore, the energy of the bound states turn out to be
\begin{equation}
|E_\pm|=\frac{8\pi^2}{m}\left(\Lambda\pm\frac{M^2}{\kappa\,g'}\right)^2,
\end{equation}
{{implying  that, for this particular choice of the Robin's boundary conditions, there exists bound states for $\kappa\in\Re$. Note also
that, for the  particular choice of $\kappa_\mp = \mp \frac{M^2}{\Lambda\,g'}$, the bound states have zero energy.}}

For scattering processes, a similar behavior is verified. Indeed, 
the problem can be formulated as follow: first we rewrite (\ref{sch12}) iteratively
\eqb
\psi ({\bf x}) &=& \psi_0 ({\bf x}) - \frac{i g'}{M^2} ~G[{\bf x},0]~ \left(\sigmavec \cdot\nablavec \psi_0 (0)\right) \nonumber 
\\
&-& \left( \sigmavec \cdot\nablavec G[{\bf x},0 ]\right)  \psi_0(0) + \cdots. \label{28}
\eqf

However, the last term in RHS of (\ref{28}) is energetically less relevant than the first one and we can write the last equation as
\bb
\label{fsc}
\psi ({\bf x}) = \psi_0 ({\bf x}) - \frac{i g'}{M^2} ~G[{\bf x},0]~ \left(\sigmavec \cdot\nablavec \psi_0 (0)\right) + \cdots. 
\ee

The appropriate Green function for the boundary conditions of the 
scattering problem  is  
$$
G[{\bf x},0] = -\frac{1}{4\pi} \frac{e^{i kr}}{r},
$$ 
and therefore scattering state (\ref{fsc}) is 
\bb 
\psi ({\bf x}) = \varphi_0 ({\bf x}) + A \frac{e^{i kr}}{r},
\ee
where, formally, 
\bb
A =-\frac{i g'}{4 \pi M^2} \left(\sigmavec \cdot\nablavec \psi_0 (0)\right)
\ee
is our definition of the scattering amplitude. 

We prepare the initial state as $\psi_0 = \bx^+\,\psi_++\bx^-\,\psi_-$ (in principle we can take $\psi_+=\psi_-=e^{\imath kz}$)
and we get
\bb 
\left(\sigmavec \cdot \nablavec \psi_0\right)_{{\bf x} =0} = -\left(\bx^-\,\partial_r\psi_+ + \bx^+\,\partial_r\psi_-\right)_{{\bf x} =0},
\ee
and, therefore, the total scattering amplitude $\sigma_{\mbox{\tiny{TOT}}} =A^\dag A$ turn out to be
\bb 
\sigma_{\mbox{\tiny{TOT}}} = \frac{{g'}^2}{16 \pi^2 M^4} \left(|\partial_r \psi_+(0)|^2 + |\partial_r \psi_-(0)|^2\right).
\ee
We impose now the Robin's boundary condition
 (\ref{cond1}) with  $\mathbb{{G}}^\dagger \mathbb{{G}}= \kappa^2 \mathbb{I} $, and therefore
total cross section  is 
\bb 
\sigma_{\mbox{\tiny{TOT}}} = \frac{{g'}^2{\kappa^2}}{16 \pi^2 M^4} \left(|\psi_+(0)|^2 + |\psi_-(0)|^2\right).
\ee

We note that, up to the normalization factor $ \left(|\psi_+(0)|^2 + |\psi_-(0)|^2\right)$, this result is  in full agreement with independent calculations  reported in \cite{gelmini} {{if we interpret $\kappa$ as the mass of the DM-nucleon}}. 



 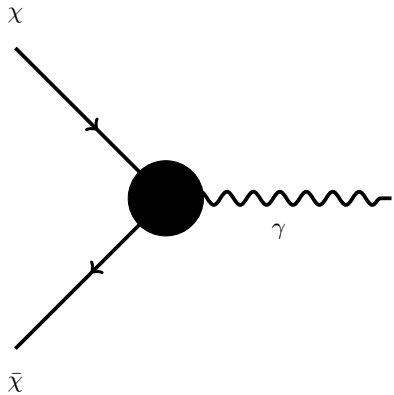
\begin{figure}
\begin{tikzpicture}[node distance=1cm and 1cm]
\coordinate (O) at (0, 0);
\draw[line width=0.5mm, aparticle] (O)--(-2,2) node[left, above=0.2cm] {$\chi$};
\draw[line width=0.5mm, particle] (O) -- (-2,-2) node[left, below=0.2cm] {$\bar{\chi}$};
\draw[line width=0.5mm, photon] (O) -- (3,0)node[midway, below=0.2cm] {$\gamma$};
\filldraw[fill=black, draw=black] (O) circle (0.5cm);
\end{tikzpicture}
\caption{This is a process of annihilation of two fermions in a photon $\gamma$. The black box is an effective vertex.}
\end{figure}
 
  From the above results we could conclude that having sufficiently reliable bounds for $\Lambda$ and $M$ could be argued about observability (or not) of  anapole dark matter. 
  
  A careful analysis of the Xenon-100 data \cite{xenon} and the bounds for hidden photons from \cite{javi} could help clarify these issues.
  \section{Final Comments}
  
  As final comments we would like to point out the following; {{ a) the anapole observables in scattering processes are only dependent on the mass scale and coupling constant that appear in the Zeldovich term and, therefore, the number of adjustable parameters needed to extract bounds  are minimal and this is  an  advantage with respect to  other approaches; b) the approach proposed here not only provides a systematic way of carrying out parity violation effects as a consequence of the standard model, but can also be considered as a starting point to study possible corrections associated with dark matter and its interactions.  }}
  
  \vskip 1.0cm 
  
  We would like to thank Manuel Asorey, Paola  Arias, and Jose G. Esteve  by discussions. This work  was  supported  by Dicyt   041831GR (J.G) and 
  041931MF (F. M.). One of us (J.G.) thanks to DESY theory group and the Alexander von Humboldt Foundation for the hospitality and support.

  \appendix
  
\section{The eigenspinors and the equation of motion}

In order to built the solution of the equation of motion, let us review some facts 
related to the spinor factorization \cite{bieden}. 

Consider the Hamiltonian
\begin{equation}
\label{h}
\cal{H}={\bf L}^2-{\bf J}^2-\frac14
\end{equation}
with the vector operators $\L={\bf r}\times{\bf p}$, the orbital angular momentum,  and $\J=\L+{\bf S}$, the 
total angular momentum. The spin operator is ${\bf S}$ and we take ${\bf S} = \frac12\,{\boldsymbol \sigma}$,
with ${\boldsymbol\sigma}$, the Pauli matrices.

It is direct to prove that 
\begin{equation}
\label{h2}
{\cal H} = - {\boldsymbol\sigma}\cdot\L-1,
\end{equation}
which will be useful in what follows. On the other hand, the eigenvalue problem 
\begin{equation}
{\cal H}\,|{\cal k}\rangle  = \kappa\,|{\cal k}\rangle
\end{equation}
has solution
\begin{equation}
\kappa =  \cdots,-2,-1,1,2,\cdots
\end{equation}
with $\kappa\neq0$.

For a given $\kappa$, the eigenvalues  of $\L^2$ and $\J^2$ ( $\ell(\ell+1)$ and $j(j+1)$, respectively) 
satisfy 
\begin{equation}
\ell = |\kappa|+\frac12\left(\mbox{sgn}(\kappa)-1\right),\quad  j = |\kappa|-\frac12.
\end{equation}

In the usual notation we have, therefore,
\begin{equation}
\label{qunum}
j=\frac12,\frac32,\frac52,\frac72\dots,
\end{equation}
with $j=\ell\pm\frac12$ and the projection $m_j$ (eigenvalue of $J_3$) are given by the $2j+1$ values
$$
m_j=-j,-j+1,-j+2,\dots,j-2,j-1,j
$$

Eigenfunctions of the commuting set of operators $\{\L^2,\J^2,J_3\}$ are the   spinor spherical harmonics (Pauli spinors)
\begin{eqnarray}
\label{plus}
{\boldsymbol \xi}_{\ell,j=\ell-\frac12}^{m_j}  &=& \left(
\begin{array}{r}
-\sqrt{\frac{\ell-m_j+\frac12}{2\ell+1}}\,Y_{\ell}^{m_j-\frac12}
\\
\,
\\
\sqrt{\frac{\ell+m_j+\frac12}{2\ell+1}}\,Y_{\ell}^{m_j+\frac12}
\end{array}\right),\quad \mbox{for }\kappa>0
\nonumber
\\
&~&
\\
\label{minus}
{\boldsymbol \xi}_{\ell,j=\ell+\frac12}^{m_j}   &=& \left(
\begin{array}{r}
\sqrt{\frac{\ell+m_j+\frac12}{2\ell+1}}\,Y_{\ell}^{m_j-\frac12}
\\
\,
\\
\sqrt{\frac{\ell-m_j+\frac12}{2\ell+1}}\,Y_{\ell}^{m_j+\frac12}
\end{array}\right), \quad \mbox{for }\kappa<0.
\nonumber
\\
\, & 
\end{eqnarray}
For a fixed total angular momentum $j$, previous spinors correspond to 
two angular momentum $\ell$, namely, $\ell =j\pm\frac12$.

Let us introduce the notation $\bx^+$ for the spinor in (\ref{plus}) and $\bx^-$ for the one in 
(\ref{minus}) omitting labels $\ell,j,m_j$. The  following relation holds
\begin{equation}
\label{actspin}
(\hat{\bf r}\cdot{\boldsymbol\sigma})\,\bx^\pm =-\bx^\mp.
\end{equation}

This relation is useful because the interaction term can be written in terms of previous operator and ${\cal H}$ only. Indeed,
from $\L=-\imath\,{\bf r}\times{\boldsymbol\nabla}$, the following identity holds
$$
{\boldsymbol\nabla} = \hat{\bf r}(\hat{\bf r}\cdot{\boldsymbol\nabla})+\frac{\imath}{r^2}{\bf r}\times\L,
$$
and therefore
\begin{equation}
{\boldsymbol\sigma}\cdot{\boldsymbol\nabla} = (\hat{\bf r}\cdot{\boldsymbol\sigma})(\hat{\bf r}\cdot{\boldsymbol\nabla})+
\frac{\imath}{r^2}{\boldsymbol\sigma}\cdot({\bf r}\times\L).
\end{equation}
The last term in previous expression can be written as
$
{\boldsymbol\sigma}\cdot({\bf r}\times\L)=\imath ({\bf r}\cdot{\boldsymbol\sigma})({\boldsymbol\sigma}\cdot\L),
$
once the relation $\sigma_i\sigma_j=\delta_{ij}+\imath\,\epsilon_{ijk}\sigma_k$ is used.

Then we get for the interaction term
$$
{\boldsymbol\sigma}\cdot{\boldsymbol\nabla} = (\hat{\bf r}\cdot{\boldsymbol\sigma})\bigg[(\hat{\bf r}\cdot{\boldsymbol\nabla})-
\frac{1}{r}({\boldsymbol\sigma}\cdot\L)\bigg].
$$

Finally, we can use (\ref{h2}) in order to replace the last term by ${\cal H}+1$. 
Since $(\hat{\bf r}\cdot{\boldsymbol\nabla}) =\partial_r$, it is possible to write the interaction term as follows
\begin{equation}
\label{inter}
{\boldsymbol\sigma}\cdot{\boldsymbol\nabla} =(\hat{\bf r}\cdot{\boldsymbol\sigma}) \bigg[\partial_r+\frac1r({\cal H
}+1)\bigg].
\end{equation}

The action of this operator on the spinors $\bx^\pm$ is obtained from (\ref{actspin})
\begin{equation}
\label{intactspin}
{\boldsymbol\sigma}\cdot{\boldsymbol\nabla} \,\bx^{\pm}=- \bx^{\mp}\,
 \bigg[\partial_r+\frac1r(1\pm|\kappa|)\bigg].
\end{equation}
Then, the interaction terms changes the orbital angular momentum $\ell$ when  the total 
angular momentum $j$ is fixed.

In particular, for a wave function with the form
$$
\psi^{\pm}({\bf x}) = \Phi^{\pm}(r)\,\bx^{\pm}(\theta,\varphi),
$$
we get
\begin{equation}
\label{finalaction}
{\boldsymbol\sigma}\cdot{\boldsymbol\nabla} \,\psi^{\pm}({\bf x}) =- \bx^{\mp}\,
 \bigg[\partial_r\Phi^{\pm}+\frac{\Phi^{\pm}}{r}(1\pm|\kappa|)\bigg].
\end{equation}

 \section{The Green function }
In this section we calculate the Green function for the Helmholtz equation.  Green function in this case
satisfies 
\begin{equation}
({\boldsymbol \nabla}^2 +k^2)G({\bf x},{\bf x}') = \delta({\bf x} -{\bf x}').
\end{equation}
%
and therefore
\begin{equation}
\label{g2}
G({\bf x}, {\bf x}') = \frac1{(2\pi)^3}\int\, \frac{\,e^{\imath {\bf p}\cdot({\bf x} -{\bf x}')}}{k^2-p^2}\,d^3p.
\end{equation}

By performing the integration on the angular variables  we get 
\begin{eqnarray}
\label{g3}
G({\bf x}, {\bf x}')
&=& \frac1{2\pi^2|{\bf x} -{\bf x}'|}\int_0^\infty\,\frac{p}{k^2-p^2}\sin(p|{\bf x} -{\bf x}'|)\,dp
\nonumber
\\
 & 
\end{eqnarray}
Last integral is divergent due to the pole at $p=k$. 

We are interested on {\it bound } states, therefore $k^2=-2m|E| \equiv-\epsilon^2$ and in such case, 
we are interested in the following expression
$$
I =\frac{1}{\Delta}\int_0^\infty\frac{p\sin(p\Delta)}{\epsilon^2+p^2}dp
$$
with $\Delta=|{\bf x} -{\bf x}'|$. The factor $\Delta^{-1}$ has been introduced for further convenience. 

In order to integrate previous expression we  introduce a cutoff  $\Lambda$ and consider, instead, 
\begin{eqnarray}
\label{il}
I_\Lambda&\equiv&\frac{1}{\Delta}\int_0^\Lambda \frac{p\sin(p\Delta)}{\epsilon^2+p^2}dp
\nonumber
\\
&=&\frac{\imath}{2\Delta}\sinh(\Delta\epsilon)\bigg[\mbox{Ci}(\Delta(\Lambda-\imath\epsilon))-
\mbox{Ci}(\Delta(\Lambda+\imath\epsilon))+\imath\pi\bigg]-
\nonumber
\\
&&~~~~\frac1\Delta\cosh(\Delta\epsilon)\bigg[\mbox{Si}(\Delta(\Lambda-\imath\epsilon))+
\mbox{Si}(\Delta(\Lambda+\imath\epsilon))\bigg]
\bigg],
\end{eqnarray}
where Si$(x)$ denotes the Sine Integral function and Ci$(x)$ is the Cosine Integral.

In order to get the behavior of this integral for $\Delta\epsilon\to 0$ and $\Delta\Lambda\to 0$
we expand in these dimensionless variables to obtain
\begin{equation}
\label{ilato}
I_\Lambda=\Lambda - \frac{\pi\epsilon}{2}+{\cal O}(\Delta^2\Lambda^2)
\end{equation}

By replacing this result in the Green function (\ref{g3}) we get the Green function in the limit ${\bf x},{\bf x}'\to 0$
\begin{eqnarray}
\label{gfin}
G_\Lambda &=& -\frac{1}{2\pi^2}\left(\Lambda+\frac{\imath \pi k}{2} \right),
\\
&=&\frac{ k}{4\pi\imath}-\Lambda
\end{eqnarray}
where we have restored $\epsilon=-\imath k$ for the bound states.

Regarding the derivative of the Green function, it is enough to  note that we are interested only in $\Delta = r$ and 
therefore, we consider  $\partial_r$.   From(\ref{il}) one obtains 
\begin{equation}
\partial_r\,I_\Lambda = \frac{\pi\Delta\epsilon^3}{6}-\frac{\Delta\epsilon^4}{3\Lambda}+{\cal O}(\Delta^3\Lambda),
\end{equation}
and therefore
$$
\partial_rG_\Lambda\to0
$$
for $r\to0$.

Finally, let us show a simpler way to obtain previous results.  In this approach, we take the limit $\Delta\to0$ before integration,
that is
\begin{eqnarray}
I_\Lambda(\Delta\to0) &=&\int_0^\Lambda\frac{p^2}{\epsilon^2+p^2}dp
\nonumber
\\
&=&\Lambda-\epsilon\,\arctan\left(\frac\Lambda\epsilon\right)
\nonumber
\\
&=&\Lambda-\frac{\pi\epsilon}{2}+{\cal O}(\epsilon/\Lambda).
\end{eqnarray}
In the limit $\Lambda\to\infty$, only the first two terms are relevant.  Finally (restoring also $\epsilon=-\imath k$) we find
$$
I_\Lambda=\Lambda+\frac{\imath\pi k}{2}
$$
which is the same result as before and therefore
$$
G_\Lambda = \frac{k}{4\pi\imath}-\Lambda.
$$

The derivative, in this approach, is direct to calculate since 
$$
\partial_\Delta\left[\frac{p\sin(p\Delta)}{\Delta(\epsilon^2+p^2)}\right] = \frac{p}{\Delta(\epsilon^2+p^2)}
\left[p\cos(p\Delta) -\frac{\sin(p\Delta)}{\Delta}\right]
$$
and it is direct to prove that in the limit $\Delta\to 0$ previous expression vanishes.

\end{document}